\documentclass[prd,twocolumn,showpacs,preprintnumbers,amsmath,amssymb]{revtex4}

\usepackage{graphicx}
\usepackage{dcolumn}
\usepackage{bm}

\newcommand{\ksl}{k \hskip -0.5em /}
\newcommand{\qsl}{q \hskip -0.5em /}
\newcommand{\psl}{P_\pi \hskip -1.0em /}

\newcommand{\bgi}{\begin{itemize}}
\newcommand{\eni}{\end{itemize}}

\newcommand{\bb}{\begin{thebibliography}}
\newcommand{\eb}{\end{thebibliography}}

\newcommand{\nn}{\nonumber \\}
\newcommand{\bwt}{\begin{widetext}}
\newcommand{\ewt}{\end{widetext}}
\newcommand{\bea}{\begin{eqnarray}}
\newcommand{\ba}{\begin{array}}
\newcommand{\ea}{\end{array}}
\newcommand{\eea}{\end{eqnarray}}
\newcommand{\be}{\begin{equation}}
\newcommand{\ee}{\end{equation}}

\newcommand{\bit}[1]{\bibitem{#1}}

\newfont{\fib}{cmfi10 at 10pt}

\newcommand{\Tr}{{\rm Tr}}

\newcommand{\N}{{\cal N}}

\newcommand{\eg}{{\it e.g.}\ }

\begin{document}

\preprint{BK-PSU HEP 02-03}

\title{A  mechanism for the $T$-odd pion fragmentation function}

\author{Leonard P.  Gamberg}
\affiliation{Division of Science,
Penn State-Berks Lehigh Valley College, 
Reading, PA 19610, USA }
\author{Gary R. Goldstein}
\affiliation{Department of Physics and Astronomy, Tufts University,
           Medford, MA 02155, USA }
\author{Karo A. Oganessyan }
\affiliation{INFN-Laboratori Nazionali di Frascati I-00044 Frascati, 
via Enrico Fermi 40, Italy
}
\affiliation{ DESY, Deutsches Elektronen Synchrotron 
Notkestrasse 85, 22603 Hamburg, Germany 
}
\date{\today}
\begin{abstract}
We consider a simple rescattering mechanism to calculate a leading 
twist $T$-odd pion  
fragmentation function, a favored candidate for filtering the transversity 
properties of the nucleon. We evaluate the single spin azimuthal asymmetry  
for a transversely polarized target in semi-inclusive deep inelastic 
scattering (for HERMES kinematics). Additionally, we calculate the double  
$T$-odd $\cos2\phi$ asymmetry in this framework.

\end{abstract}

\pacs{12.38.-t, 13.60.-r, 13.88.+e}
\maketitle

\section{\label{intro} Introduction}
\vskip-0.25cm
The transversity distribution, $h_1$ (also known as $\delta q$), which 
measures the probability to find a transversely polarized quark in the 
transversely polarized nucleon, is as important for the description 
of the internal nucleon structure and its spin properties as the more familiar 
longitudinal distribution function, $g_1$. However, it still remains 
unmeasured, unlike the spin-average and helicity distribution 
functions, which are known experimentally and extensively modeled 
theoretically. The difficulty is that $h_1$ is a chiral-odd function, 
and consequently suppressed in 
inclusive deep inelastic scattering (DIS) processes~\cite{jaffe91}; it has 
to be accompanied by a second chiral-odd quantity. In semi-inclusive deep 
inelastic scattering (SIDIS) of 
transversely polarized nucleons several methods have been proposed to access
transversity distributions. The more promising one relies on 
the so called Collins fragmentation function~\cite{cnpb93}, which  
correlates the transverse spin of the fragmenting quark to the transverse 
momentum of the produced hadron. Beside being chiral-odd, this fragmentation 
function is also time-reversal odd ($T$-odd, see \eg~\cite{boer,ANSL}), 
which makes its 
calculation challenging.  Earlier, in addition to the Collins 
parameterization~\cite{cnpb93}, a 
theoretical attempt was made to estimate  the Collins function for 
pions~\cite{artru1}. More recently, a 
non-vanishing $T$ -odd fragmentation function was obtained through a 
consistent  one-loop calculation, where massive 
constituent quarks and pions are the only 
effective degrees of freedom~\cite{BKMM}.  In addition to parameterizations
from data indicating a 
non-zero Collins function~\cite{efre},
the  non-zero single spin asymmetries in  recent  
measuriments~\cite{HERMES,SMC,STAR}
signal the existence of a non-trivial $T$-odd effects.
These indications of the $T$-odd fragmentation functions
 taken together call for deeper investigations,  
both  theoretical and
experimental. 

In this paper we explore an alternative  one-gluon exchange 
mechanism, for the fragmentation 
of a transversely polarized quark into a spinless
hadron similar to the approach we 
applied~\cite{gamb_gold_ogan1,gamb_gold_ogan2} 
to the distribution of the transversely polarized quarks in 
the both unpolarized and transversely polarized nucleon
(in this context see also~\cite{bhs,ji,ji2}). Within this consistent
framework we now make predictions of the leading 
twist $T$-odd pion fragmentation function, and the resulting
 single transverse spin azimuthal asymmetry, $\sin(\phi+\phi_S)$, 
as well as the spin-independent, $\cos 2\phi$, asymmetry in SIDIS.

\section{\label{model} The Mechanism}
\vskip-0.25cm

The non-perturbative information about the quark content of the target 
and the fragmentation of quarks into hadrons 
in SIDIS is encoded in 
the general form of the factorized cross sections in terms of  
the quark distributions $\Phi(p)$ and  fragmentation functions $\Delta(k)$, 
 entering the hadronic tensor
\bea
M{\cal W}^{\mu\nu}(P,P_h,q)&=&\int d^4 k d^4 p\delta^4(k+q-p)
\nn &&\hspace{-1.75cm}
\times 
{\rm Tr}\left(\gamma^\mu\Phi(k)\gamma^\nu\Delta(p)\right)
+\left(
\begin{array}{ccc} 
q & \leftrightarrow  & -q \\ 
\mu & \leftrightarrow  & \nu
\end{array}
\right),
\eea
to leading 
order in $1/Q^2$~\cite{mulders2}.  
Here $k$ and $p$ are the quark scattering and 
fragmenting momenta and $P$ and $P_h$ are the target and
fragmented hadron momenta respectively. Further, $\Phi$ is given by
\bea
\Phi(p,P)&=&\frac{1}{2}\int dp^- \Phi(p,P)
\Big|_{p^+=xP^+}
\nn&=&
\sum_X
\int {\frac{d\xi^- d^2\xi_\perp }
  {(2\pi)^3}} e^{ip\cdot \xi}
\langle P|\overline{\psi}(\xi^-,\xi_\perp)
{\cal G}^{\dagger}_{{\scriptscriptstyle{[\xi^-,\infty]}}}
\big|X\rangle
\nn  && \hspace{1cm} 
\langle X\big|
{\cal G}_{{\scriptscriptstyle{[0,\infty]}}}
\psi(0)|P\rangle\vert_{{\scriptscriptstyle{\xi^+=0}}}.
\label{DF}
\eea
The fragmentation matrix is, 
\bea
\Delta(k,P_h) &=&\frac{1}{4z}\int dk^+ \Delta(k,P_h)
\Big|_{k^-=\frac{P_h}{z}}
\nn &=& 
\sum_X \int \frac{d\xi^+d^2 \xi_\perp}{2z\,(2\pi)^3} \ 
e^{ik\cdot \xi} \,\langle 0\ \vert 
{\cal G}_{{\scriptscriptstyle{[\xi^+,-\infty]}}} \psi (\xi) 
\vert X;P_h\rangle
\nn && \hspace{1cm} 
\left.
\langle X;P_h\vert \overline \psi(0) 
{\cal G}^\dagger_{{\scriptscriptstyle{[0,-\infty]}}}
\vert 0 \rangle \right|_{\xi^- = 0},
\label{FF} 
\eea
where the path ordered exponential along the light like 
direction $\xi^-$ is 
\bea
{\cal G}_{[\xi^-,\infty]}={\cal P}
\exp{\left(-ig\int_{\xi^-}^\infty d\xi^- A^+(\xi)\right)},
\label{link}
\eea
and $\{\big|X\rangle \}$ represents a complete set of states.

The path ordered light-cone link operator is necessary to maintain 
gauge invariance and appears to respect factorization~\cite{cplb,ji,ji2}. 
Further, 
in non-singular gauges~\cite{ji,ji2}, Eqs.~(\ref{DF},\ref{FF})  give rise
to initial and final state interactions which in turn 
provide a mechanism to generate leading  twist $T$-odd 
contributions to both the distribution and {\em fragmentation}
functions.  The joint product of these functions enter
novel azimuthal asymmetries and single spin asymmetries (SSAs) 
that have been reported in the 
literature~\cite{ji,ji2,bhs,gold_gamb,bbh,gamb_gold_ogan1,gamb_gold_ogan2}.
These functions have been defined via the 
transverse momentum dependent quark distributions,
implicit in Eqs.~(\ref{DF},\ref{FF}),
by employing the identities for manipulating the limits of
the ordered exponential 
and the Feynman rules for such processes~\cite{col82,cplb}.

 Such an analysis was recently applied to the $T$-odd 
($f_{1T}^\perp$)~\cite{ji,sivers} 
and $h_1^\perp$~\cite{gold_gamb,gamb_gold_ogan1,bbh,gamb_gold_ogan2} 
distribution functions
(in addition to baryon fragmentation functions~\cite{metz}).
Here we will apply an analogous procedure to generate 
the $T$-odd pion fragmentation function, $H_1^\perp(z)$. 
In turn we will analyze both 
the $\cos2\phi$ and the $\sin(\phi+\phi_S)$ asymmetries in semi-inclusive 
pion electroproduction.

\section{\label{col}
Pion Fragmentation Function}

\begin{figure}
\includegraphics[height=4.0 cm]{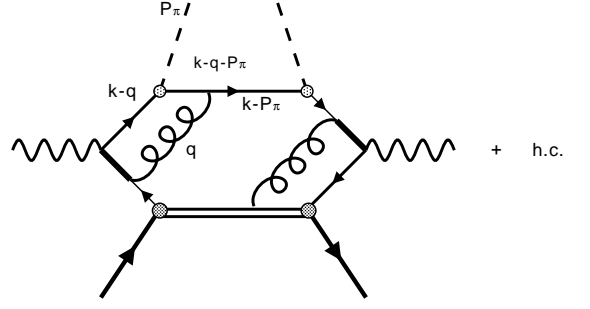}
\caption{\label{eik} Figure depicts $h_1^\perp\star H_1^\perp$ $\cos 2\phi$ 
asymmetry. The momenta flow to the quark-pion vertex is shown. 
The momentum $q$ is the loop integration variable. The $T$-odd distribution
and fragmentation functions in this approach are obtained from cutting  the
spectators.}
\end{figure}

With the tree level contribution vanishing for $T$-odd functions, 
the leading order  contributions 
to the (see~\cite{gamb_gold_ogan2} for the quark distribution) 
$T$-odd 
fragmentation functions come from the first non-trivial term in expanding
 the path ordered gauge link operator in Eq.~(\ref{link})
\bea
\Delta(z,k_\perp)&=&{\frac{1} {4z}}\sum_X
\int {\frac{d\xi^- d^2\xi_\perp }
  {(2\pi)^3}} e^{i(\xi^- k^+-\vec{\xi}_\perp \vec{k}_\perp)} 
\nn && \hspace{-1cm}
\langle 0|\left(ig\int^\infty_0 d\xi^+ A^-(\xi^+,0) 
   \right){\psi}(\xi^-,\xi_\perp)|P_h;X\rangle
\nn && 
\langle P_h; X|
\overline{\psi}(0,0_\perp)|0\rangle\vert_{\scriptscriptstyle{\xi^-=0}}  
\quad +\quad {\rm h. c.}\, .
\nn
\label{FF1}
\eea
The corresponding Feynman rules are those 
for interactions between an eikonalized struck
quark and the remaining target depicted in Fig.~\ref{eik}.  
In modeling the highly off-shell
fragmenting quark we adopt a minimal spectator~\cite{mel,hood,jac} approach. 
Here,  we couple the on-shell spectator, as a quark  
interacting with the produced pion (hereafter, $P_h=P_\pi$)
through the vertex function
\be
\langle 0\big| \psi(0)\big| P;X\rangle= 
\left(\frac{i}{\ksl
 - m}\right)\Upsilon(k_\perp^2) u({\scriptstyle k-P_\pi},s),
\ee
where 
\bea
\Upsilon(k_\perp^2)=i \gamma_5 f_{qq\pi}{\N^\prime}
e^{-b^\prime k^2_{\perp}} .
\eea
We have introduced a Gaussian distribution
in the transverse momentum dependence of the quark-spectator-pion 
vertex ~\cite{hood,gamb_gold_ogan2} in order to address
the $\log$ divergence
that arises from the moments of fragmentation functions. 
Here, $f_{qq\pi}$ (defined henceforth as $f$) is the quark-pion
coupling and $k$ is the momentum of the off-shell quark, 
$k_\perp$ and $b^\prime=1/<k_\perp^2>$, are the  intrinsic 
transverse momentum and its inverse mean square, respectively, 
${\N^\prime}$ is a dimensionful normalization, 
and finally, $u(p,s)$,  is the on-shell quark spinor.
The  Collins function, is projected from Eq.~(\ref{FF1}) 
\bea
\Delta^{\scriptscriptstyle [\sigma^{\perp +}\gamma_5]}(z,k_\perp)
=\frac{1}{4z}\int dk^+ \Tr\left(\gamma^-\gamma^\perp\gamma_5\Delta\right)
\Big|_{k^-=\frac{P_\pi}{z}}.
\label{delta}
\eea
The leading order (in $1/Q$) one loop contribution to Eq.~(\ref{delta}), 
which arises in the limit that the $-$ 
light-cone component of the virtual photon's
momentum goes to infinity~(see \eg~\cite{ji2}), 
corresponding to the rescattering
of the initial state quark depicted  in Fig.~\ref{eik},
is given by the expression
\bea
\Delta(k,P_\pi)\hspace{-0.1cm}&=&\hspace{-0.25cm}\int
\frac{d^4q}{\left(2\pi\right)^4}\frac{\ksl-\qsl+m}{(k-q)^2-m^2}
\nn && \hspace{-1.5cm}
\frac{\N^\prime f\gamma_5\left(\ksl-(\qsl+{\psl}\ )+\mu\right)}
{\left(k-q-P_{\pi} \right)^2-\mu^2}\frac{g^2\gamma^-}{q^-+i\epsilon}
\frac{1}{q^2-\lambda_g^2+i\epsilon}
\nn && \hspace{-1.5cm}
\delta((k-P_\pi)^2-\mu^2)(\ksl-{\psl}\ +\mu)
\frac{\N^\prime f\gamma_5\left(\ksl+m\right)}{k^2-m^2},
\nn
\eea
where we have taken the sum over states to be saturated by a single 
spectator quark with effective mass $\mu$. The off-shell fragmenting quark's 
momentum is given by~\cite{berger,hood}
\bea
k^2=\frac{zk^2_\perp}{1-z}+\frac{\mu^2}{1-z}+\frac{M_\pi^2}{z},
\label{virt}
\eea
which results in the correct counting rules for scalar meson
production in limit $z\rightarrow 1$.
The details of the loop integration
are similar to those performed in~\cite{gamb_gold_ogan1,gamb_gold_ogan2}.  
We evaluate the projection
$\Delta^{\scriptscriptstyle [i\sigma^{\perp +}\gamma_5]}$, which    
results in the
leading twist, $T$-odd  contribution
\bea
H_1^\perp(z,k_\perp) 
&=&\frac{{\N^\prime}^2 f^2g^2}{(2\pi)^4}\frac{1}{4z}\frac{(1-z)}{z}
\frac{m}{\Lambda^\prime(k^2_\perp)}
\frac{M_\pi}{k_\perp^2}
\nn && \hspace{-.75cm}
e^{-b^\prime\left(k^2_\perp- \Lambda^\prime(0)\right)}
\left[\Gamma(0,b\Lambda^\prime(0))\hspace{-.10cm}-\hspace{-.10cm}
\Gamma(0,b^\prime\Lambda^\prime(k^2_\perp))\right] ,
\nn
\eea
where,
$\Lambda^\prime(k^2_\perp)=k_\perp^2 +\frac{1-z}{z^2}M_\pi^2
+ \frac{\mu^2}{z} -\frac{1-z}{z} m^2$. 
The average $<k^2_{\perp}>$ or $b^\prime$ is a regulating scale 
which we fit to the expression for the integrated unpolarized 
fragmentation function
\bea
D_1(z)&=&\frac{{\N^\prime}^2 f_{qq\pi}^2}{4(2\pi)^2}\frac{1}{z}
\frac{\left(1-z\right)}{z}
\Bigg\{\frac{m^2-\Lambda^\prime(0)}{\Lambda^\prime(0)}
\nn && \hspace{-.75cm}
-\left[2b^\prime\left(m^2-\Lambda^\prime(0)\right)-1\right]
e^{2b^\prime\Lambda^\prime(0)}\Gamma(0,2b^\prime\Lambda^\prime(0))\Bigg\},
\nn
\eea
which, multiplied by $z$ at $<k_\perp^2> = {(0.5)}^2$ GeV$^2$ and $\mu=m$,
is in  good agreement with the
distribution of Ref.~\cite{kretz}. It is important to emphasize
that the kinematics of the virtuality condition of the decaying
quark Eq.~(\ref{virt}) enforces
the proper dimensional counting rules~\cite{berger,hood}.
\begin{figure}
\includegraphics[height=7.0 cm]{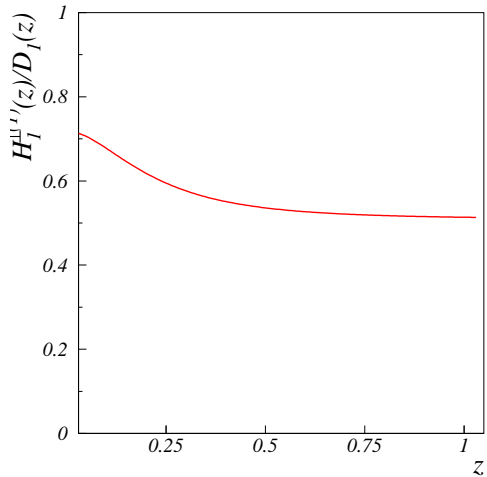}
\caption{\label{analyze} The weighted analyzing power
$H_1^{\perp (1)}(z)/D_1(z)$ as a function of $z$.}
\end{figure}

The chiral odd transversity distribution,
$h_1$, for a scalar spectator diquark in the quark-diquark model of the 
nucleon is
\bea
h_1(x)&=&\frac{g^2{\N}^2}{4(2\pi)^2}\left(1-x\right)
\nn && \times
\left(m+xM\right)^2\Bigg\{\frac{1}{\Lambda(0)}-2be^{2b\Lambda(0)}
\Gamma(0,2b\Lambda(0))\Bigg\}.
\nn
\eea
and the previously calculated~\cite{gamb_gold_ogan2} 
unpolarized distribution is
\bea
f(x)&=&\frac{g^2{\N}^2}{4(2\pi)^2}\left(1-x\right)
\Bigg\{\frac{\left(m+xM\right)^2-\Lambda(0)}{\Lambda(0)}
\nn &&\hspace{-1.25cm}
-\left[2b\left(\left(m+xM\right)^2-\Lambda(0)\right)-1\right]
\times e^{2b\Lambda(0)}\Gamma(0,2b\Lambda(0))\Bigg\}\, ,
\nn
\eea
where $g$ contains the gluon-scalar diquark coupling, and 
$\Lambda(0)=(1-x)m^2 +x\lambda^2  -x(1-x)M^2$, while 
$M$, $m$, and $\lambda$ are the nucleon, quark, and diquark masses 
respectively.  Choosing $<p_\perp^2> = {(0.4)}^2$ GeV$^2$ = $1/b$  
yields  good agreement with the valence
distribution of Ref.~\cite{GRV}.

We consider this to be a reasonable phenomenological framework, 
which avoids  the log divergence~\cite{hood,gamb_gold_ogan2} 
involved in integrating over $k_\perp$ and $p_\perp$, 
while introducing an average transverse momentum determined from spin 
averaged scattering~\cite{ellis,chay}.  Additionally, this 
form factor approach is compatible with the
parameterization of the fragmentation functions employed 
in references~\cite{kotz,OABD}
to set the Gaussian width for the fragmentation function.
In Fig.~\ref{analyze} the weighted 
the analyzing power, $H_1^{\perp (1)}(z)/ D_1(z)$, is displayed. 
The resulting behavior is similar to a previous model ansatz 
proposed by Collins and calculated in Ref.~\cite{kotz}.

\section{\label{ASSA} \bf Azimuthal asymmetries}
\vskip-0.25cm

We discuss the explicit result and numerical evaluation of the
single transverse-spin $\sin (\phi + \phi_s)$ and double $T$-odd 
$\cos2\phi$ asymmetries for $\pi^+$ production in SIDIS.
\begin{figure}
\includegraphics[height=6.5 cm]{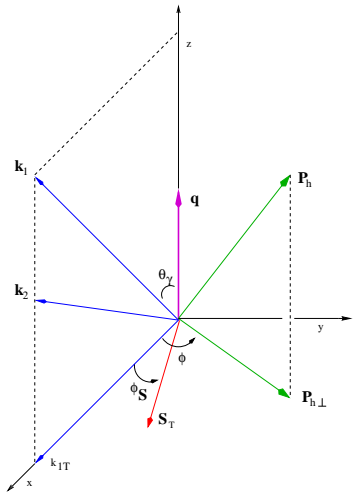}
\caption{\label{kin}The kinematics of semi-inclusive DIS: $k_1$ ($k_2$) is 
the 4-momentum of the 
incoming (outgoing) charged lepton,  where $q=k_1-k_2$, 
is the 4-momentum of the virtual photon. $P$ ($P_h$) is 
the momentum of the target (observed) hadron. The scaling
variables are $x=Q^2/2P\cdot q$ ,  $y=P\cdot q/P\cdot k_1$ , 
and $z=P\cdot P_h/P\cdot q$. 
The momentum $k_{1T}$ ($P_{h\perp}$) is the incoming lepton (observed hadron) 
momentum component perpendicular to the virtual photon momentum direction. 
$\phi_s$ and $\phi$ are the azimuthal angles, 
of the target spin projection $S_{\mathbf T}$
and  $P_{h\perp}$  respectively.}
\end{figure}

The $\cos2\phi$ asymmetry of SIDIS is projected out of the cross section 
and depends on a leading double $T$-odd product, 
\bea
{\langle \frac{\vert P^2_{h{\perp}} \vert}{M M_\pi} \cos2\phi \rangle}
{\scriptscriptstyle_{UU}}&=& 
\frac{\int d^2P_{h\perp} \frac{\vert P^2_{h\perp}\vert}{M M_\pi}
\cos 2\phi\,  d\sigma}
{\int d^2 P_{h\perp}\, d\sigma} 
\nn &=&\frac{{8(1-y)} \sum_q e^2_q h^{\perp(1)}_1(x) z^2 H^{\perp(1)}_1(z)}
{{(1+{(1-y)}^2)}  \sum_q e^2_q f_1(x) D_1(z)}
\label{ASY_cos} 
\nn
\eea
where the subscript $UU$ indicates unpolarized beam and 
target (Note: The non-vanishing 
$\cos2\phi$ asymmetry originating from the $T$-even distribution and 
fragmentation function 
appears only at order $1/Q^2$~\cite{CAHN,kotz0,OBDN}).
$h_1^{\perp (1)}(z)$ is the weighted
moment of the distribution  
function~\cite{boer,gamb_gold_ogan2}.
The SIDIS differential cross section depends on variables $x, y, z$ 
and azimuthal angles $\phi$ and $\phi_s$ (see Fig.~\ref{kin}). 
For a transversely polarized target 
nucleon, the $\sin(\phi+\phi_s)$ asymmetry\cite{cnpb93,mulders2} 
can be projected 
out with an 
azimuthal integration, yielding, the convolution of two chiral-odd (both 
$T$-odd and $T$-even) structures,
\bea
\langle \frac{P_{h\perp}}{M_\pi}
\sin(\phi+\phi_s) \rangle_{\scriptscriptstyle UT}
\hspace{-0.2cm}&&
\nn && \hspace{-2.25cm}
=\frac{\int \hspace{-0.1cm} d\phi_s \int \hspace{-0.1cm}
{ d^2P_{h\perp}}\frac{P_{h\perp}}{M_\pi}
\sin(\phi\hspace{-0.1cm}+\hspace{-0.1cm}\phi_s)
\left(d\sigma^{\uparrow}\hspace{-0.1cm}-\hspace{-0.1cm}
d\sigma^{\downarrow}\right)}
{\int d\phi_s\int d^2P_{h\perp}\left(d\sigma^{\uparrow}
+d\sigma^{\downarrow}\right)}
\nn && \hspace{-2.25cm}
=\big|S_T\big|\frac{2(1-y) \sum_q e^2_q h_1(x) z H^{\perp(1)}_1(z)}
{{(1+{(1-y)}^2)}  \sum_q e^2_q f_1(x) D_1(z)}.
\nn
\label{ASY_sin}
\eea
\begin{figure}
\includegraphics[height=7.0 cm]{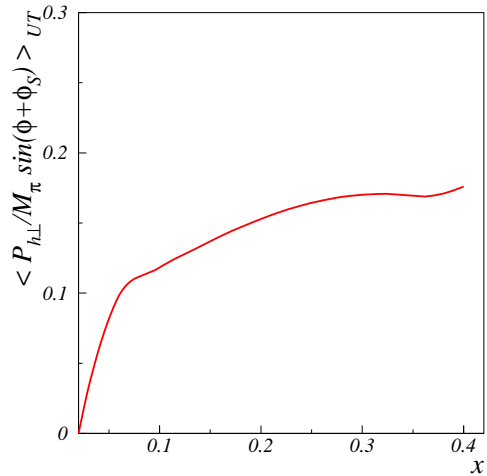}
\includegraphics[height=7.0 cm]{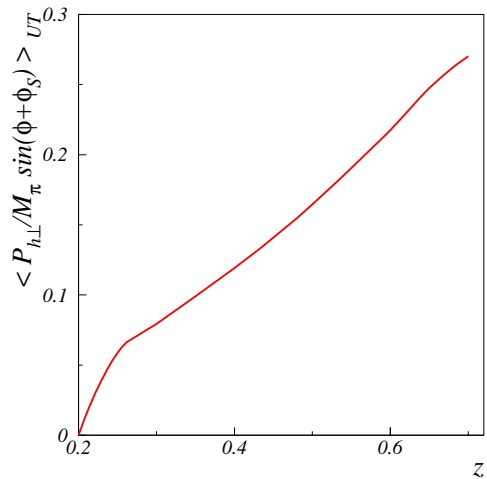}
\caption{\label{collins}Upper Panel: The 
\protect{${\langle \frac{P_{h\perp}}{M_h} \sin(\phi+\phi_s) 
\rangle}_{\scriptscriptstyle UT}$ } asymmetry for $\pi^+$ production as 
a function of 
 $x$ .  
Lower Panel: The 
\protect{$ {\langle \frac{P_{h\perp}}{M_h} \sin(\phi+\phi_s) \rangle }
_{\scriptscriptstyle UT}$} 
asymmetry for $\pi^+$ production as a function of  $z$.}
\end{figure}
\begin{figure}
\includegraphics[height=7.0 cm]{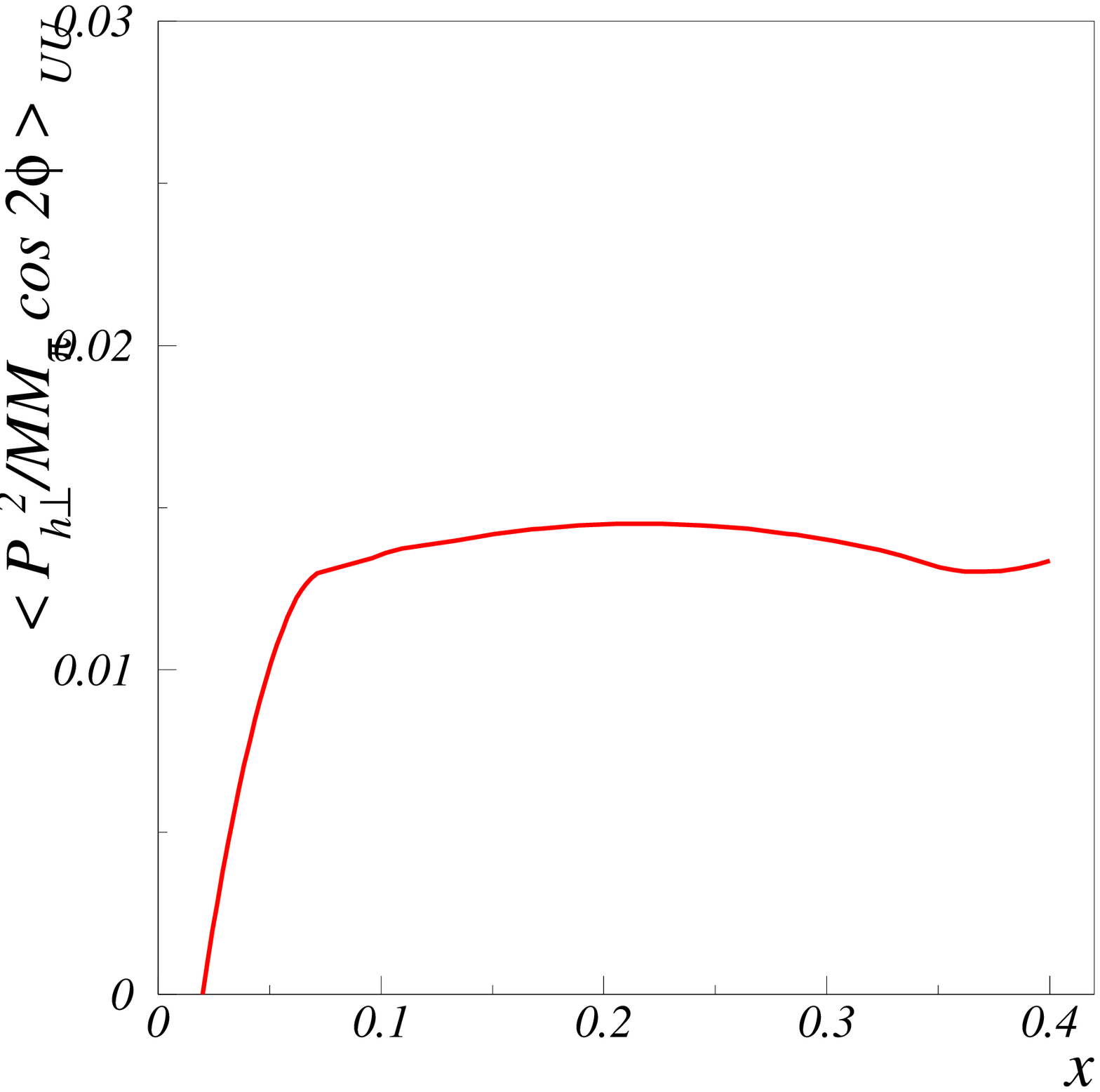}
\includegraphics[height=7.0 cm]{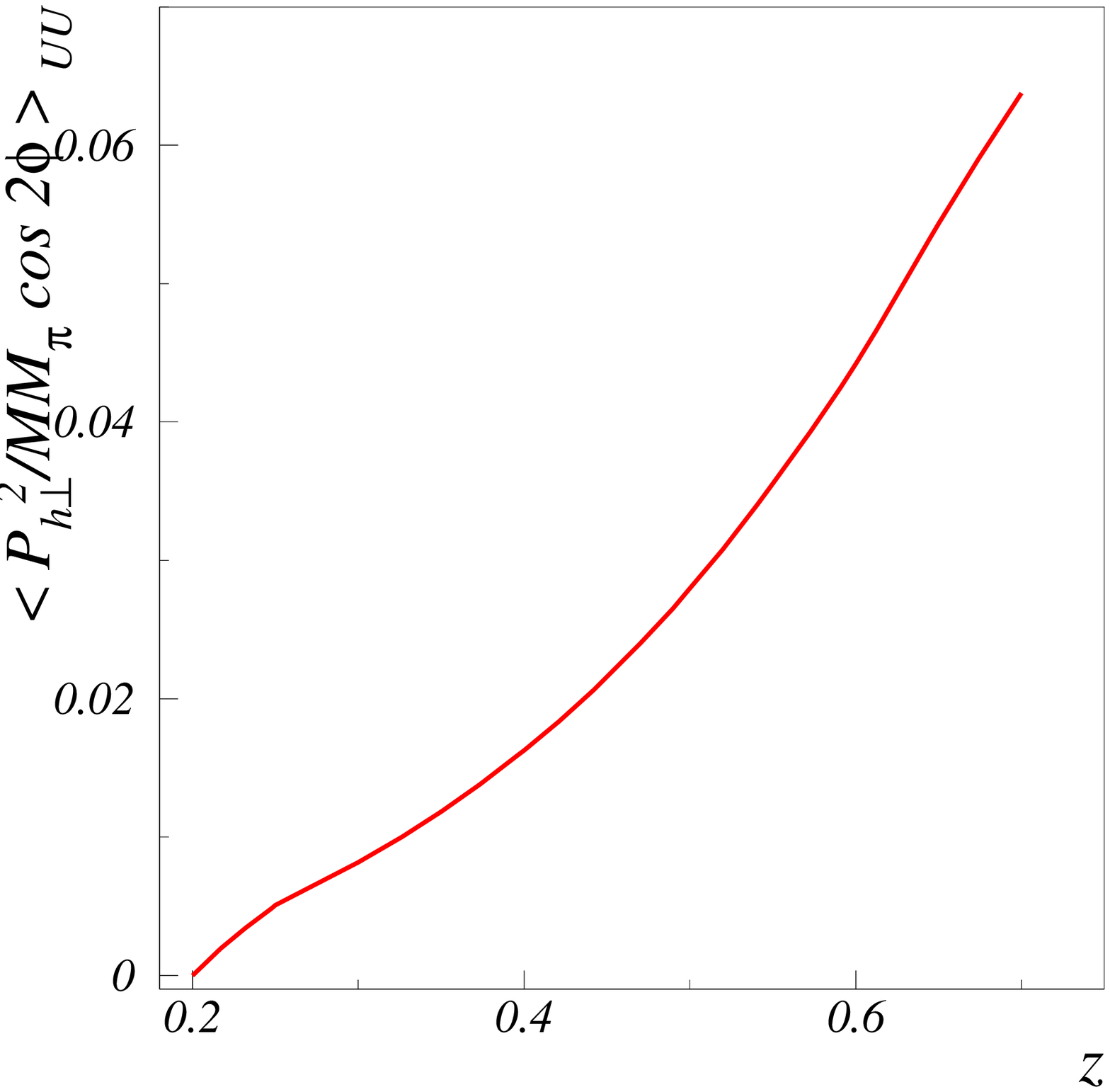}
\caption{\label{cos2}Upper Panel: The \protect{${\langle 
\frac{P^2_{h\perp}}{MM_h} \cos2 \phi \rangle}_{\scriptscriptstyle UU}$} 
asymmetry for \protect{$\pi^+$} production 
as a function of \protect{$x$}. 
Lower Panel: The 
\protect{${\langle \frac{P^2_{h\perp}}{MM_h} \cos2\phi 
\rangle}_{\scriptscriptstyle UU}$} asymmetry for 
\protect{$\pi^+$} production 
as a function of \protect{$z$}.}
\end{figure}

We define the variable range to coincide with the HERMES kinematics,
$1$ GeV$^2$ $\leq Q^2 \leq 15$ GeV$^2$, $4.5$ GeV 
$\leq E_{\pi} \leq 13.5$ GeV, $0.2 \leq z \leq 0.7$, $0.2 \leq y \leq 0.8$.  
In Fig.~\ref{collins} the 
${ \langle \frac{P_{h\perp}}{M_{\pi}} \sin(\phi+\phi_s) 
\rangle}_{\scriptscriptstyle UT}$ 
asymmetry of Eq.~(\ref{ASY_sin}) 
for $\pi^+$ production on a proton target is presented 
as a function of $x$ and $z$, respectively (using $\Lambda_{QCD}=0.2$ GeV) 
Fig.~\ref{collins} indicates approximately a  
$10-15\%$  $P_{h\perp}/M_{\pi}$ weighted $\sin(\phi+\phi_s)$  asymmetry.
Similarly, in Fig.~\ref{cos2} the  $P^2_{h\perp}/(M M_{\pi})$ weighted 
$\cos 2\phi$ asymmetry of Eqs.~(\ref{ASY_cos}) 
for $\pi^+$ production on an unpolarized proton target is presented as 
a function of $x$ and $z$, respectively. Fig.~\ref{cos2} indicates 
a few percent asymmetry.  

\section{Conclusion}
\vskip-0.2cm
A mechanism to generate the $T$-odd Collins 
fragmentation function that is derived from the gauge link
has been considered. This approach complements and is 
consistent with the approach that was employed to 
generate the Sivers $f_{1T}^\perp$
 and the chiral odd unpolarized $h_1^\perp$ distribution functions
that fuel the Sivers and $\cos 2\phi$ asymmetries. In order to
consistently calculate these asymmetries it is advantageous 
to generate the Collins fragmentation function from this framework.
The derivation of the Collins function 
is consistent with the observation that intrinsic transverse quark momenta and 
angular momentum conservation are intimately tied with studies of 
transversity~\cite{kotz0}. 
This was demonstrated previously from analyzes of the tensor charge in the 
context
of the axial-vector dominance approach to exclusive meson 
photo-production~\cite{gamb_gold}, and to SSAs in 
SIDIS~\cite{gold_gamb,gamb_gold_ogan1,gamb_gold_ogan2}. 
Furthermore, this approach is interesting in that it does not suffer from
the possible cancellation of the Collins effect cited in~\cite{jin_jaffe};
namely, that phases from final state interactions of the pions
with the spectator remnant will sum to zero. 
This mechanism does not 
rely on multiple interactions with the outgoing pion. 
On the contrary, the effect is generated in the 
non-trivial phase associated with the gauge link 
operator~\cite{ji,cplb,ji2,metz,bbh,gamb_gold_ogan1,gamb_gold_ogan2,pij}.

In  the case of unpolarized beam and target, we have predicted that at HERMES
energies there is a non-trivial $\cos2\phi$ asymmetry associated with the
asymmetric distributions of transversely polarized quarks inside
unpolarized hadrons. We have evaluated the
analyzing power and predicted
the $P_{h\perp}/M_{\pi}$ weighted $\sin(\phi+\phi_S)$ asymmetry.
We note also that the analyzing power, $H_1^{\perp (1)}(z)/D_1(z)$
displays behavior characteristic of previous results where other
methods were used to characterize the Collins mechanism.
This ratio is consistent 
with the Collins ansatz~\cite{cnpb93,kotz}. 
Generalizing from these model calculations, it is clear that initial and 
final state  interactions can account for  leading twist 
$T$-odd contributions to SSAs.  
In addition, while it has been shown that other 
mechanisms, ranging from  
initial state interactions to the non-trivial phases  of 
light-cone wave functions~\cite{bhm,bhs,metz}, 
can give rise to SSAs, these various mechanisms can be understood in the
context of gauge fixing as it impacts the gauge link operator
in the transverse momentum quark distribution 
functions~\cite{ji,ji2,gamb_gold_ogan1,gamb_gold_ogan2}.
Thus, using rescattering as a mechanism to generate 
$T$-odd distribution and fragmentation functions 
opens a new window into the theory and phenomenology of transversity in 
hard processes. 

Note added in proof: After our manuscript was submitted for publication
a paper appeared on a similar subject~\cite{bac_metz}.
\begin{acknowledgments} 
LPG is supported in part by funds provided from a
Research Development Grant, Penn State Berks, and GRG from 
the US Department of Energy,   {\small DE-FG02-29ER40702}.
\end{acknowledgments}

\bibliography{apssamp}

\bb{99}

\bibitem{jaffe91}J. Ralston and D. E. Soper, Nucl. Phys. {\bf B152},109 (1979);
X. Artru and M. Mekhfi, Z. Phys. C {\bf 45}, 669 (1990) ;
R. L. Jaffe and X. Ji, Phys. Rev. Lett. {\bf 67}, 552 (1991) ;
Nucl. Phys. {\bf B375}, 527 (1992).

\bibitem{cnpb93} J.C. Collins, Nucl. Phys. {\bf B396},161 (1993).

\bibitem{boer} D. Boer and P.J. Mulders, Phys. Rev. D {\bf 57}, 5780 (1998).

\bibitem{ANSL} M. Anselmino and F. Murgia, Phys. Lett B {\bf 442}, 470 (1998).

\bibitem{artru1} X. Artru , J. Czyzewski, and  H. Yabuki 
 Z.Phys.C {\bf 73}, 527 (1997).

\bibitem{BKMM} A. Bacchetta, R. Kundu, A. Metz, and P.J. Mulders 
Phys. Lett. B{\bf 506} 155( 2001); Phys.Rev.D {\bf 65} 094021 (2002).

\bibitem{efre} A.V. Efremov, K. Goeke, and P. Schweitzer, 
Phys.Lett.B {\bf 522}, 37 (2001).

\bibitem{HERMES} A. Airapetian {\it et al.}, Phys. Rev. Lett.
 {\bf 84}, 4047 (2000); Phys. Rev. D {\bf 64}, 097101 (2001); 
Phys. Lett. B {\bf 562}, 182 (2003).

\bibitem{SMC}  A. Bravar (Spin Muon Collaboration), Nucl. Phys. Proc. 
Suppl., {\bf 79} 520 (1999). 

\bibitem{STAR} L.C. Bland, hep-ex/0212013; G. Rakness, hep-ex/0211068.  

\bibitem{gamb_gold_ogan1} L. P. Gamberg, G. R. Goldstein and 
K.A.~Oganessyan, hep-ph/0211155, To be Published in the {\em Proceedings
of the $15^{\rm th}$ International Spin Physics Symposium (SPIN 2002)}, 
Long Island, New York, September 2002.

\bibitem{gamb_gold_ogan2} L. P. Gamberg, G. R. Goldstein and 
K.A.~Oganessyan, Phys. Rev. D {\bf 67 }, 071504 (2003).

\bibitem{bhs} S. Brodsky, D.S. Hwang and I. Schmidt, 
Phys. Lett. B {\bf 530}, 99 (2002).

\bibitem{ji} X. Ji and F. Yuan, Phys. Lett. B {\bf 543}, 66 (2002).

\bibitem{ji2} A.V. Belitsky, X. Ji and F. Yuan, hep-ph/0208038.

\bibitem{mulders2} R.D. Tangerman and P. J. Mulders,
Phys. Lett. B {\bf 352}, 129 (1995) ; Phys. Rev. D {\bf 51}, 3357 (1995);
Nucl. Phys. {\bf B461}, 197 (1996).

\bibitem{cplb} J.C. Collins, Phys. Lett. B {\bf 536}, 43 (2002).

\bibitem{gold_gamb} G. R. Goldstein and L. P. Gamberg, hep-ph/0209085,
To be published
in the {\em Proceedings of $31^{\rm st}$ International Conference on 
High Energy Physics (ICHEP 2002)}, Amsterdam, The Netherlands, Jul 2002. 

\bibitem{bbh} D. Boer, S. Brodsky, D.S. Hwang, 
Phys. Rev D {\bf 67}, 054003  (2003).

\bibitem{col82} J. C. Collins and D. E, Soper, Nucl. Phys. {\bf B194},
445 (1982); J. C. Collins, D. E, Soper and G. Sterman in
{\em Perturbative Quantum Chromodynamics} ed. A. H. Mueller (World
Scientific, 1989), p. 1.

\bit{sivers} D. Sivers, Phys. Rev D 41 (1990) 83 ; Phys. Rev. D {\bf 43}, 261 
(1991).

\bibitem{metz} A. Metz, Phys. Lett. B {\bf 549}, 139 (2002).

\bibitem{mel} W. Melnitchouk , A.W. Schreiber, and A.W. Thomas, 
Phys. Rev.D {\bf 49}, 1183 (1994).
 
\bibitem{hood} M. Nzar and 
P. Hoodbhoy, Phys.Rev. D {\bf 51}, 32 (1995).

\bibitem{jac} R. Jakob, P.J. Mulders and J. Rodriques, Nucl. Phys.
{\bf A626}, 937 (1997); A. Bacchetta, S. Boffi, and R. Jakob, 
Eur. Phys. J {\bf A9}, 131 (2000).

\bibitem{berger} E. L. Berger, Phys. Lett B {\bf 89}, 241 (1980).

\bibitem{kretz} S. Kretzer, Phys. Rev. D {\bf 62}, 054001 (2000).

\bibitem{GRV} M.~Gl\" uck, E.~Reya, and A.~Vogt, Z. Phys. C {\bf 67}, 433 
(1995).

\bibitem{ellis} R.K. Ellis, W.J. Stirling and B.R. Webber, {\it QCD and
Collider Physics} (Cambridge University Press, Cambridge, U.K. 1996),
p.305.

\bibitem{chay} J. Chay, S. D. Ellis, and  W.J. Stirling, 
Phys.Rev. D {\bf 45},46 (1992).

\bibitem{kotz} A. M. Kotzinian and P. J. Mulders,
Phys. Lett. B {\bf 406}, 373 (1997).

\bibitem{OABD} K.A.~Oganessyan,  H. R. Avakian,  N. Bianchi and P. Di Nezza, 
Eur. Phys. J.  C {\bf 5}, 681 (1998).

\bibitem{kotz0} A. M. Kotzinian, Nucl. Phys {\bf B441}, 234 (1995).
\bibitem{CAHN} R.N. Cahn, Phys. Lett.  B {\bf 78}, 269 (1978) ; 
Phys. Rev. D {\bf 40}, 3107 (1989). 
\bibitem{OBDN} K.A.~Oganessyan, N. Bianchi, E. De Sanctis, and 
W.D. Nowak, Nucl. Phys. {\bf A689}, 784 (2001).

\bibitem{gamb_gold} L. P. Gamberg and G. R. Goldstein, Phys.
Rev. Lett. {\bf 87}, 242001 (2001).

\bibitem{jin_jaffe} R. L. Jaffe, X. Jin, and J. Tang, 
Phys. Rev. Lett {\bf 80}, 1166 (1998).

\bibitem{pij} D.  Boer, P.J. Mulders, and F. Pijlman,  hep-ph/0303034.

\bibitem{bhm} S. J. Brodsky, P. Hoyer, N. Marchal, S. Peign\'{e}, and
F. Sannino, Phys. Rev. D {\bf 65}, 114025 (2002).

\bibitem{bac_metz} A. Bacchetta, A. Metz, and J.J. Yang, hep-ph/0307282.

\eb

\end{document}